\documentclass[journal ]{new-aiaa}
\usepackage[utf8]{inputenc}
\usepackage{textcomp}
\usepackage{graphicx}
\usepackage{amsmath}
\usepackage[version=4]{mhchem}
\usepackage{siunitx}
\usepackage{longtable,tabularx}
\usepackage{svg}
\title{Performance of Oscillating Plasma Thrusters}

\author{Jacob Simmonds\footnote{PhD Candidate, Mechanical and Aerospace Engineering}}
\affil{Princeton University, Princeton, New Jersey 08543}
\author{Yevgeny Raitses\footnote{Principal Research Investigator, Plasma Science and Technology}}
\affil{Princeton Plasma Physics Laboratory, Princeton, New Jersey 08543}

\begin{document}

\maketitle

\begin{abstract}
Traditional expressions and definitions describing performance of plasma thrusters, including the thrust, specific impulse, and the thruster efficiency, assume a steady state plasma flow with a constant flow velocity. However, it is very common for these thrusters that the plasma exhibits unstable behavior resulting in time-variations of the thrust and the exhaust velocity. For example, in Hall thrusters, the ionization instability leads to strong oscillations of the discharge current (so-called breathing oscillations), plasma density, ion energy, and as a result the ion flow. In this paper, we revisit the formulation of the thrust and the thrust efficiency to account for time variations of the ion parameters including the phase shift between the ion energy and the ion flow. For sinusoidal oscillations it was found thrust can potentially change more than 20\%. It is shown that by modulating ion energy at specific amplitudes, thrust can be maximized in such regimes. Finally, an expression for the thruster efficiency of the modulating thruster is derived to show a mechanism for inefficiencies in such thrusters.

\end{abstract}

Submitted to \emph{Journal of Propulsion and Power} Dec. 28, 2019

\section*{Nomenclature}
{\renewcommand\arraystretch{1.0}
\noindent\begin{longtable*}{@{}l @{\quad=\quad} l@{}}
$T$  & thrust, N\\
$\dot{m}$ &   mass flow, kg/s \\
$v_{jet}$& jet velocity, m/s \\
$v_{ex}$& exhaust velocity, m/s \\
$P_{in}$ & input electrical power, W \\
$P_{thrust}$ & thrust power, W \\
$P_{Kinetic}$ & kinetic power, W \\
${\eta}$   & total efficiency \\
${\eta_{curr}}$ & current efficiency \\
${\eta_{volt}}$ & voltage efficiency \\
${\eta_{prop}}$ & propellant utilization \\
${\eta_{osc}}$ & oscillation factor \\
${\eta_{ipr}}$ & ion power ratio \\
$V$ & voltage, V \\
$I$ & current, A \\
$\phi$  & phase difference, rad \\
$M$  & mass of ion, kg \\
$e$  & charge, C\\
$Isp$  & specific impulse, s\\
\multicolumn{2}{@{}l}{Subscripts}\\
$i$ & ion\\
$d$ & discharge\\
$n$ & neutral\\
$m$ & mean\\
$a$ & amplitude\\
$tot$ & total
\end{longtable*}}

\section{\label{sec:level1}Introduction}

For many plasma thrusters, plasma instabilities often result in unstable thruster operation which may affect thruster performance. For example, operation of Hall thrusters often exhibits discharge current oscillations of 10-30 kHz. These discharge oscillations result in oscillations of the plasma flow from the thruster \cite{BouefLowFreq,ParkerOscillations,BarralOscillations}. Apart from these naturally occurring oscillations, recent studies have explored externally driven oscillations of the discharge to control the thruster operation \cite{RomadanovDriven,HaraDrivenBreathing,ShiDriven,YamamatoPuls}. In particular, work by Romadanov et al. modulated the DC discharge voltage with a sinusoidal signal. It was found that thrust under such oscillating regimes may degrade when the ion flow and ion energy oscillations shift out of phase. In addition, these experiments revealed that this phase shift becomes appreciable, particularly at large oscillation amplitudes \cite{SimmondsMod}. A key question then is what effect this phase shift may have on thruster performance, including the thrust, Isp and thruster efficiency. To the best of our knowledge this question has never been addressed in plasma thruster literature. Therefore, the main goal of this paper is to derive and compare time-resolved and time-averaged thrust and thruster efficiency expressions and relative values for oscillating plasma thrusters.

This paper is organized as follows. Section 2 discusses performance for a steady-state plasma thruster operation and shows the importance of accounting for time-dependent oscillations in the formulation of the thruster performance. Section 3 derives thrust for a sinusoidal modulated thruster. Section 4 introduces the expressions for the input power under modulation and the performance of modulated thrusters. Finally ion power ratio in the oscillating plasma thruster is discussed in Section 5.

\section{From Steady-State to Time-Dependent}
\subsection{Remarks on Steady State Thruster Performance}

For a steady-state operation of the thruster, the thrust is defined as

\begin{equation} \label{eq:ThrustNoModsec}
{T} = {\dot{m}_{tot}} {v_{jet}}
\end{equation}
where $\dot{m}_{tot}$ is the total mass flow and $v_{jet}$ is the jet velocity. For plasma thrusters with input power $P_{in}$, the efficiency is defined as the ratio between the thrust power $P_{thrust}$ and the input electrical power:

\begin{equation} \label{eq:EfficiencyNoModsec}
{\eta} = \frac{P_{thrust}}{P_{in}} = \frac{T^2}{2\dot{m}_{tot}P_{in}}
\end{equation}
\begin{equation} \label{eq:Power}
{P_{in}} = I_{d}V_{d}
\end{equation}
where $I_{d}$ and $V_{d}$ are discharge current and voltage respectively. When the plasma flow and the plasma exhaust velocity oscillate about a mean value, the derivation of the thrust becomes much more involved. For the sake of our analysis, we consider the time-dependent thrust produced by ion acceleration in an applied electric field. This mechanism of ion acceleration is relevant to Hall thrusters and ion thrusters. Detailed descriptions of how the electric field is generated can be found in other references \cite{SanchezOverview}. The time-dependence of the thrust is due to either natural plasma oscillations or external modulations of the applied power or voltage.

\subsection{Transient Time Scales}
\begin{figure}
\centering
\includegraphics[width=0.45\textwidth]{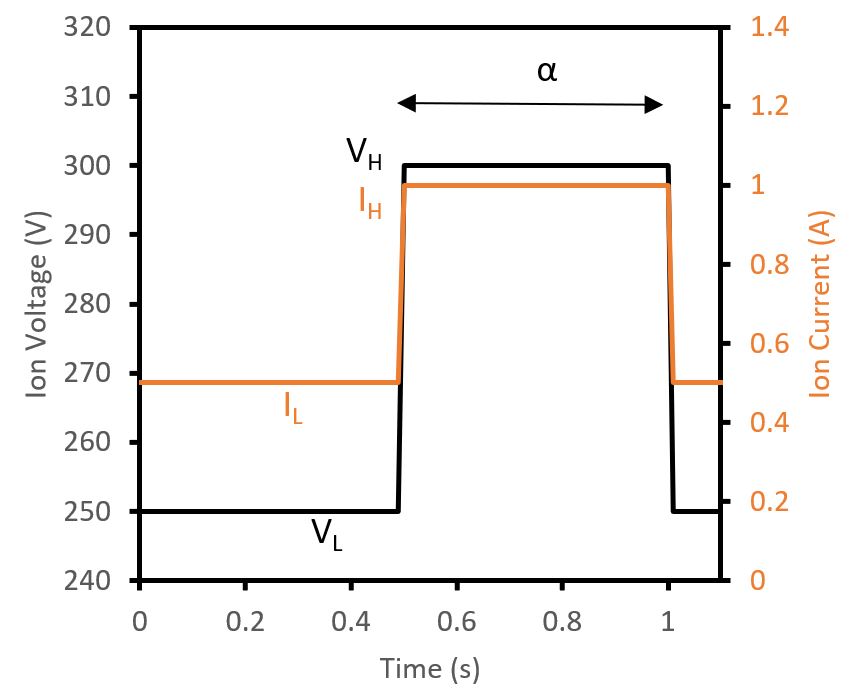}
\caption{Ion Voltage and Ion Current with Time at 1Hz Pulsed Operation}
\label{fig:lowfreqtimes}
\end{figure}

Consider a plasma thruster operating in a pulsed mode where the frequency of pulse repetition is much slower than the time scale of the transient plasma processes, such as a breathing instability oscillating the ion current. For such regimes, the derivation of the time-dependent performance is relatively simple. Due to the negligible timescales of the transient processes, the accelerating voltage (energy/charge) and mass flow are in phase (Fig. \ref{fig:lowfreqtimes}) and so the thrust is dependent only on the duty cycle ($\alpha$) of the pulsing. Pulsed thrust can be expressed as
\begin{equation} \label{eq:ThrustLowFreq}
\begin{aligned}
{T_{pulsed}} &= \frac{1}{\tau} \int_{0}^{\tau} \dot{m}(t) v_{ex}(t) dt\\
&=  \sqrt{\frac{2M}{e}}\left(I_{L} \sqrt{V_{L}}(1-\alpha)+ I_{H} \sqrt{V_{H}}\alpha \right)
\end{aligned}
\end{equation}
where $I_L$ and $I_H$ are the low and high levels of the ion current respectively, $V_L$ and $V_H$ are the low and high levels of the ion energy respectively, and $\alpha$ is the duty cycle of the square-wave pulse.

Consider now that the repetition frequency increases to approach the transient time-scales of the thruster (e.g. the frequency of natural oscillations or the pulse duration). Under such conditions, non-linear thruster plasma response (e.g. resonance-kind behavior \cite{RomadanovDriven} or a hysteresis) can cause the ion current to lag or lead the ion energy, and in some cases, alter the shape of the waveforms. Limiting the current analysis to square waves, the performance should depend on both the duty cycles of each and their phase shift $\phi_{i}$. This causes the relatively simple square pulse case to be considerably more difficult as there can be multiple solutions depending on the relative duty cycles and phases of the ion energy and ion current flow. Appendix A contains the exact solutions of the thrust for a high-frequency square wave oscillation for each of these 6 cases.

While it is not difficult to derive, the sheer variety of solutions for thrust for the various square-wave forms is inconvenient for usage. Therefore, for the rest of the paper, a sine wave oscillation is considered for two reasons. The first is that there is a single solution to the time-dependent thrust and power, and the second is this solution is directly relevant to the situation explored in experiments with externally-driven breathing oscillations in Hall thrusters \cite{RomadanovDriven,DialloTimeResolve,YamamatoPuls}. A comparison of theoretical thrust and power with experiments will be a subject of a separate paper.

\section{Modulated Thrust}

\begin{figure}
\centering
\includegraphics[width=0.45\textwidth]{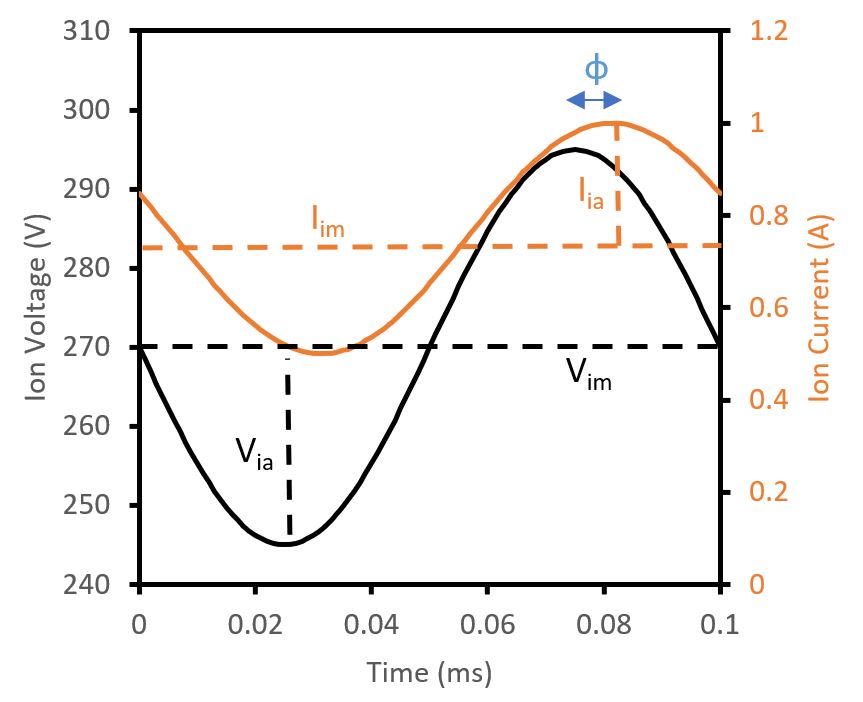}
\caption{Ion Voltage and Ion Current with Time at 10kHz Sine-wave Operation}
\label{fig:highfreqsine}
\end{figure}

For simplicity of our analysis, we assume the oscillations in ion current and the energy are sinusoidal and of the form (Fig. \ref{fig:highfreqsine}):

\begin{equation} \label{eq:CurrentEnergyOsc}
\begin{gathered}
    {I_i(\theta)} = I_{im} + I_{ia}\sin{(\theta + \phi_{i})}\\
    {V_i(\theta)} = V_{im} + V_{ia}\sin{\theta}
\end{gathered}
\end{equation}
Here $I_{im}$ is mean ion current, $I_{ia}$ is amplitude of ion current oscillations, $V_{im}$ is mean ion energy, and $V_{ia}$ is amplitude of ion energy oscillations, and $\phi_{i}$ is the phase angle between current and ion energy. Ion energy is expressed as the accelerating voltage. Thrust can be derived by finding the time average of the product of the instantaneous ion mass flow and energy, as shown in Eq. \ref{eq:ThrustMod1sec}.

\begin{equation} \label{eq:ThrustMod1sec}
 {T_{mod}} = \frac{1}{\tau} \int_{0}^{\tau} \dot{m}(t) v_{ex}(t) dt
\end{equation}
Substituting our equations for ion current and ion energy for mass flow and $v_{ex}$ respectively, this expression can be solved into a form with elliptic integrals of the first and second kind. The full derivation for thrust is shown in Appendix B and comes out to

\begin{equation} \label{eq:ThrustModFinalSec}
\begin{gathered}
 T_{mod} =  \sqrt{\frac{2 M V_{im}}{e}}\left( I_{im}\left(1- \sum_{n=1}^{\infty} a_{2n}\,  \bar \upsilon^{2n}\right)\right.\\
  + \left.I_{ia}\cos\phi_{i}\sum_{n=1}^{\infty} a_{(2n-1)}\,\bar \upsilon^{(2n-1)}\right)
  \end{gathered}
\end{equation}
where $\bar\upsilon = V_{ia}/V_{im}$ and the coefficients $a_n$ can be found in Appendix D. A low-error approximation of this expression for practical usage including associated error can also be found in Appendix B. 

\begin{figure}
\centering
\includegraphics[width=0.45\textwidth]{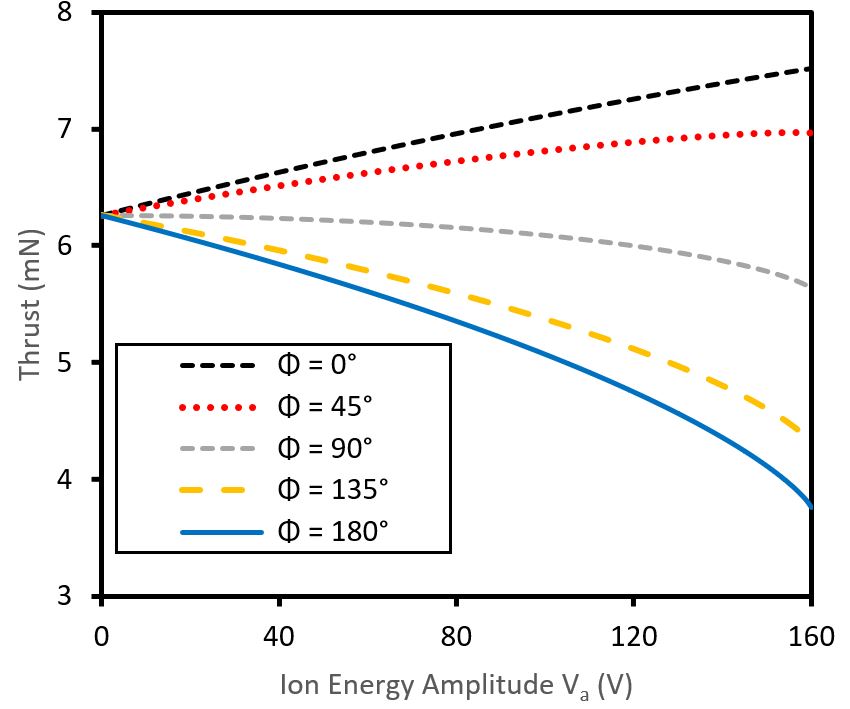}
\caption{Calculated thrust vs energy amplitude over several phase angles. Thruster parameters: $V_{im} = 160 V$, $I_{im}=I_{ia}=0.3 A$ with sinusoidal modulation}
\label{fig:phaseenergydependence}
\end{figure}

The modulated thrust in Eq. \ref{eq:ThrustModFinalSec} can be separated into the sum of three terms: the steady state thrust of the mean voltage and current, a portion that decreases thrust, and a portion that increases thrust:

\begin{equation} \label{eq:ThrustModFinalSec2}
T_{mod} =  T_{steady}- T_{D} + T_{I}
\end{equation}

\begin{equation} \label{eq:ThrustModSteady}
T_{steady} = \sqrt{\frac{2 M V_{im}}{e}} I_{im} = {\dot{m}_{tot}} {v_{jet}}
\end{equation}

\begin{equation} \label{eq:ThrustModDecrease}
T_{D} =  \sqrt{\frac{2 M V_{im}}{e}} I_{im} \sum_{n=1}^{\infty} a_{2n}\, \bar \upsilon^{2n}
\end{equation}

\begin{equation} \label{eq:ThrustModIncrease}
 T_{I} =  \sqrt{\frac{2 M V_{im}}{e}} I_{ia}\cos\phi_{i}\sum_{n=1}^{\infty} a_{(2n-1)}\,\bar \upsilon^{(2n-1)}
\end{equation}
Unintuitive effects of oscillation on thrust can be seen in Eq. \ref{eq:ThrustModFinalSec2} and illustrated in Fig. \ref{fig:phaseenergydependence}. Here plots of thrust vs ion energy amplitude are shown using typical ion energy and ion currents found in recent modulation experiments with a cylindrical Hall thruster (CHT): $I_{im}=0.3 A$, $I_{ia}=0.3 A$, and $V_{im}=160V$ \cite{SimmondsMod}. Larger modulations in ion energy $\bar \upsilon$ lower the thrust through $T_{D}$, but these same modulations increase the $T_{I}$ portion, scaled by $\cos \phi_{i}$. This results in a net increase in thrust at low phase angles, but a decrease at high phase.

\begin{figure}
\centering
\includegraphics[width=0.45\textwidth]{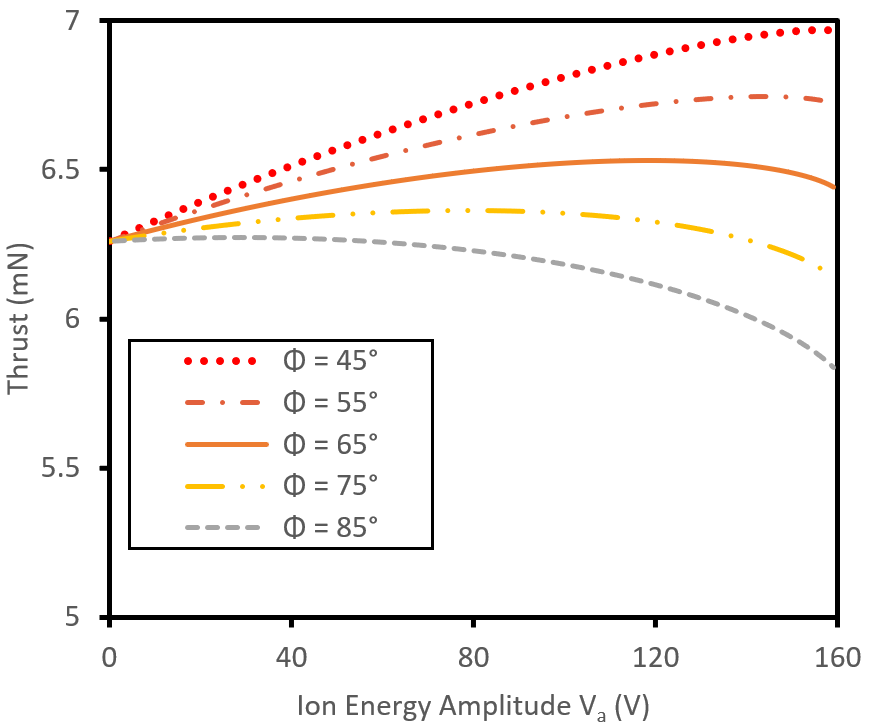}
\caption{Calculated thrust vs energy amplitude from 45$^{o}$ to 85$^{o}$. Thruster parameters: $V_{im} = 160 V$, $I_{im}=I_{ia}=0.3 A$ with sinusoidal modulation}
\label{fig:phaseenergydependencemaxes}
\end{figure}

A large phase angle implies a higher portion of ions are accelerated at a lower voltage and contribute less to thrust. As the voltage oscillation increases this effect gets worse, which can lower the thrust by as much as 40\% below the DC level when phase is 180$^{o}$. At low phase angles a higher portion of ions are accelerated at high voltage and so thrust increases. The transition region between these two effects holds some interest as the "boost" or decrease in thrust is not linear with oscillation amplitude. At mid-range phase angles (between roughly 45$^{o}$ to 85$^{o}$) the thrust initially increases with ion energy amplitude before decreasing. This causes a maximum in thrust which can be seen in Fig. \ref{fig:phaseenergydependencemaxes}. This nonlinearity in the thrust is dependent on the modulating waveform, and it suggests a theoretical maximum of the thrust for the thruster with modulated operation or oscillating thruster. Taking the derivative of the thrust equation (Eq. \ref{eq:ThrustModFinalSec}) with respect to $\bar \upsilon$, it is possible to find an oscillating regime that would theoretically provide the maximum thrust for a given phase angle and ion current. Solving for the maximum thrust $V_{ia}$ involves finding the roots of nth degree polynomials, depending on the order of expansion as shown in Eq. \ref{eq:maxthrustder2} in Appendix C.

\begin{equation} \label{eq:maxthrustder1}
\frac{I_{ia}\cos{\phi_{i}}}{I_{im}} = \frac{\sum_{n=1}^{\infty} 2n\,a_{2n}\, \upsilon^{2n-1}}{\sum_{n=1}^{\infty} (2n-1)\,a_{(2n-1)}\,\upsilon^{(2n-2)}}
\end{equation}

\begin{figure}
\centering
\includegraphics[width=0.45\textwidth]{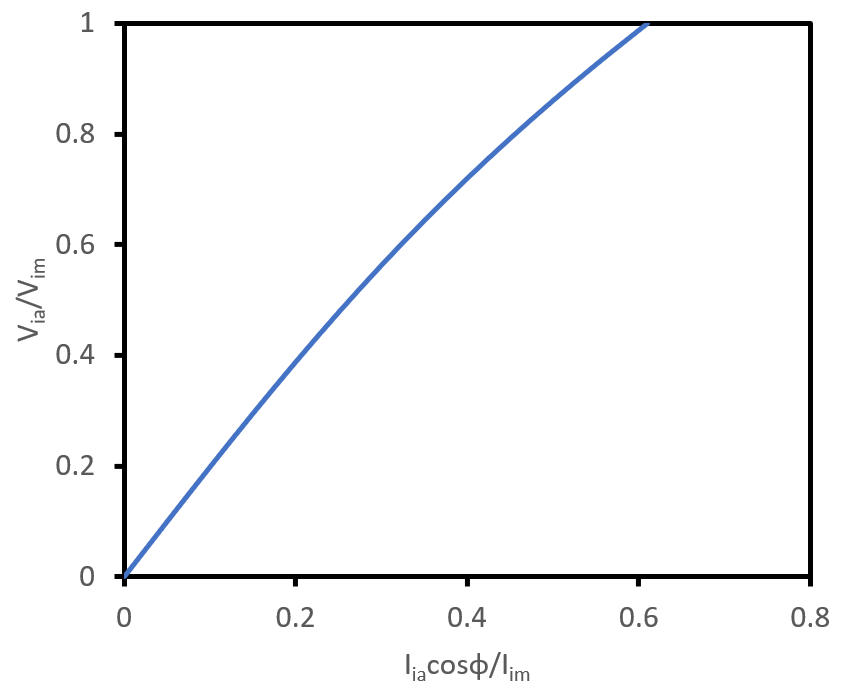}
\caption{$V_{ia}/V_{im}$ that provides maximum thrust with respect to current flow parameter $I_{ia}\cos\phi_{i}/I_{im}$ for sinusoidal modulation}
\label{fig:maxthrustva}
\end{figure}

Hence, the dimensionless $\bar \upsilon$ which provides the maximum thrust depends solely on another dimensionless parameter: $I_{ia}\cos{\phi_{i}}/I_{im}$. Solving the relationship between the two dimensionless parameters can be done numerically. The curve on Fig. \ref{fig:maxthrustva} corresponds to the optimal relationship between the dimensionless parameters at which a theoretical thrust maximum can be achieved for the thruster with oscillations. This theoretical maximum thrust is only applicable in a small transition regime where the term $I_{ia}\cos{\phi_{i}}/I_{im}$ is not too large. Thus it does not present the highest theoretically possible thrust for the sinusoidal waveform.

\section{Performance}
\subsection{Input Power} \label{powersection}
To find the efficiency of the oscillating plasma thruster, in addition to the thrust (Eq. \ref{eq:ThrustModFinalSec2}), it is also necessary to account for effects of oscillations on the input power. For example, in Hall thruster experiments with externally driven oscillations of the discharge voltage, a phase difference between the discharge current and discharge voltage was often observed \cite{YamamatoPuls}. Thus the expression for the input power of oscillating thruster has to be different from the expression used for conventional DC power Hall and Ion thrusters.

Consider the case of a sinusoidally oscillating input voltage offset by some DC level, much like the form of ion energy and ion current.
\begin{equation} \label{eq:InputCurrentVoltage}
\begin{gathered}
    {I_d(\theta)} = I_{dm} + I_{da}\sin{(\theta + \phi_{d})}\\
    {V_d(\theta)} = V_{dm} + V_{da}\sin{\theta}
\end{gathered}
\end{equation}
where $I_{dm}$ is mean discharge current, $I_{da}$ is amplitude of discharge current oscillations, $V_{dm}$ is mean discharge voltage, and $V_{da}$ is the amplitude of discharge voltage oscillations, and $\phi_{d}$ is the phase angle between discharge current and voltage.
The input power can be found by integrating the product of the discharge current and voltage which comes out to
\begin{equation} \label{eq:InputPowerDerive}
\begin{aligned}
{P_{in}} &=
    \frac{1}{2 \pi} \int_{0}^{2 \pi} (I_{dm} + I_{da}\sin{(\theta + \phi_{d})})(V_{dm} + V_{da}\sin{\theta}) d\theta\\
    & = I_{dm} V_{dm}  + I_{da} V_{da}\cos{(\phi_{d})}/2
\end{aligned}
\end{equation}
The power can then be seen as mean component ($I_{id} V_{dm}$) of the power plus the AC component $I_{da} V_{da}\cos{(\phi_{d})}/2$, which is dependent on the phase angle $\phi_{d}$. The input electric power will be at minimum when current and voltage are out of phase and is at maximum when the two are in phase.

\subsection{Efficiency}\label{sec:performance}

Using Equation \ref{eq:EfficiencyNoModsec}, the thruster efficiency is often split into three separate components: current efficiency, voltage efficiency, and propellant utilization \cite{HoferEfficiency}. Other terms such as plume divergence will not be considered here. Taking the typical current efficiency ($\eta_{curr} = I_{im} / I_{dm}$),  voltage efficiency ($\eta_{volt} = V_{im} / V_{dm}$), and propellant utilization ($\eta_{prop} = \dot{m}_i/(\dot{m}_i+\dot{m}_n)$), efficiency is

\begin{equation} \label{eq:TotalEfficiency1}
{\eta} =  \eta_{curr} \eta_{volt} \eta_{prop}
\end{equation}
This is a useful form of efficiency and desirable to keep. The difficulty in doing so arises from the fact that not only are ion energy and ion current oscillating in time, the discharge voltage and current may be too. The product of the mean of each of these is not equal to the mean of the product of these terms. The time-dependent effects, such as phasing differences, may alter the efficiency. However it is possible to derive an efficiency that maintains this form with the addition of another term which includes these oscillatory effects. Starting by deriving thrust power $P_{Thrust}=T^2/(2\dot{m}_{tot})$ and continuing the derivation form of thrust described in Appendix B, where the terms A and B contain elliptic integrals:

\begin{equation} \label{eq:ThrustPowerMod}
\begin{aligned}
{P_{Thrust}} &= \frac{\left(\overline{T}\right)^2}{2 {\dot{\overline{m}}_{tot}}}\\
&= \frac{\left( \frac{1}{\tau} \int_{0}^{\tau} \dot{m}(t) v_{ex}(t) dt\right)^2}{2 \left( \frac{1}{\tau} \int_{0}^{\tau} \dot{m}(t) dt\right)}\\
&= \frac{(A + B)^2 }{4 \pi^2 I_{im}}
\end{aligned}
\end{equation}
The last equation is specific for a plasma thruster with sinusoidal oscillations of ion current and ion energy. Substituting  Eq. \ref{eq:ThrustPowerMod} and Eq. \ref{eq:InputPowerDerive} into Eq. \ref{eq:EfficiencyNoModsec}, assuming only singly charged ions ($\eta_{prop}=1$), and expanding thrust to the second order:
\begin{equation} \label{eq:TotalEfficiency2}
\begin{aligned}
{\eta} &= \frac{P_{Thrust}}{P_{in}}\\
&= \frac{\frac{(A + B)^2 }{4 \pi^2 I_{im}}}{I_{dm} V_{dm}  + I_{da} V_{da}\cos{(\phi_{d})}/2}\\
&= \frac{I_{im}}{I_{dm}}\frac{ V_{im}}{V_{dm}}  \frac{ \left(1-\frac{\bar \upsilon^2}{16}+\bar i\,\bar \upsilon\cos{(\phi_{i})}/4\right)^2 }{(1+\bar i_d\,\bar \upsilon_d\cos(\phi_{d})/2)}\\
&= \eta_{curr} \eta_{volt} \frac{ \left(1-\frac{\bar \upsilon^2}{16}+\bar i\,\bar \upsilon\cos{(\phi_{i})}/4\right)^2 }{(1+\bar i_d\,\bar \upsilon_d\cos(\phi_{d})/2)}\\
&= \eta_{curr} \eta_{volt} \eta_{osc}
\end{aligned}
\end{equation}
We achieve a form similar to the typical efficiency equation where $\bar i = I_{ia}/I_{im}$, $\bar i_d = I_{da}/I_{dm}$, and $\bar \upsilon_d = V_{da}/V_{dm}$. This oscillation term $\eta_{osc}$ accounts for the phase variations both in the discharge power and ion thrust power.
\begin{equation} \label{eq:effpowerutil}
\eta_{osc} = \frac{ \left(1-\frac{\bar \upsilon^2}{16}+\bar i\,\bar \upsilon\cos{(\phi_{i})}/4\right)^2 }{(1+\bar i_d\,\bar \upsilon_d\cos(\phi_{d})/2)}
\end{equation}

From the above equation it can be seen that for an oscillating thruster the performance is dependent on both the ion phase $\phi_{i}$ and the discharge phase $\phi_{d}$. The form of Eq. \ref{eq:effpowerutil} is specific to a sinusoidally modulating thruster with thrust expanded to the second order, but a similar equation may be found for further orders of expansion or for different waveforms. Note that because Eq. \ref{eq:TotalEfficiency1} contains propellant utilization, the above derivations for the efficiency (Eq. \ref{eq:TotalEfficiency2} and Eq. \ref{eq:effpowerutil}) were done assuming full ionization of the propellant, i.e. no neutral species. If there is a neutral species with a much lower velocity than the ion species, the propellant utilization term can be introduced as follows, following the typical derivation for propellant utilization \cite{HoferEfficiency},

\begin{equation} \label{eq:effproputiltotal}
\eta = \eta_{curr} \eta_{volt} \eta_{osc} \eta_{prop} 
\end{equation}

The numerator of the oscillatory $\eta_{osc}$ term alters the performance due to the alignment of the ion velocity with the ion flow. What is particularly interesting is the decrease in performance due to higher amplitude of ion energy (and ion velocity) by the second term of the numerator, even when the ion energy is in phase with the ion flow. This term occurs due to the difference between thrust power and the kinetic power of the ions in the thruster exhaust, which requires some analysis.

\section{Ion Power Ratio}\label{sec:powerutilsection}

It can be shown that when the ion velocity distribution function (VDF) is described by a delta function (no energy spread), the thrust power is equal to the kinetic power of the accelerated ion flow. However, when the ion VDF is broader than the delta function, the thrust power is no longer equal to the kinetic power. Given some random VDF of the ions, the thrust power is proportional to the squared mean of the particles’ velocity (jet velocity), while the kinetic power is proportional to the squared quadratic mean (root-mean-square) of the velocity. This is evident when considering the form of thrust power and kinetic power of the exhaust. Consider some velocity distribution $v$ with associated mass flow $\dot{m}$:

\begin{figure}[h]
\centering
\includegraphics[width=0.45\textwidth]{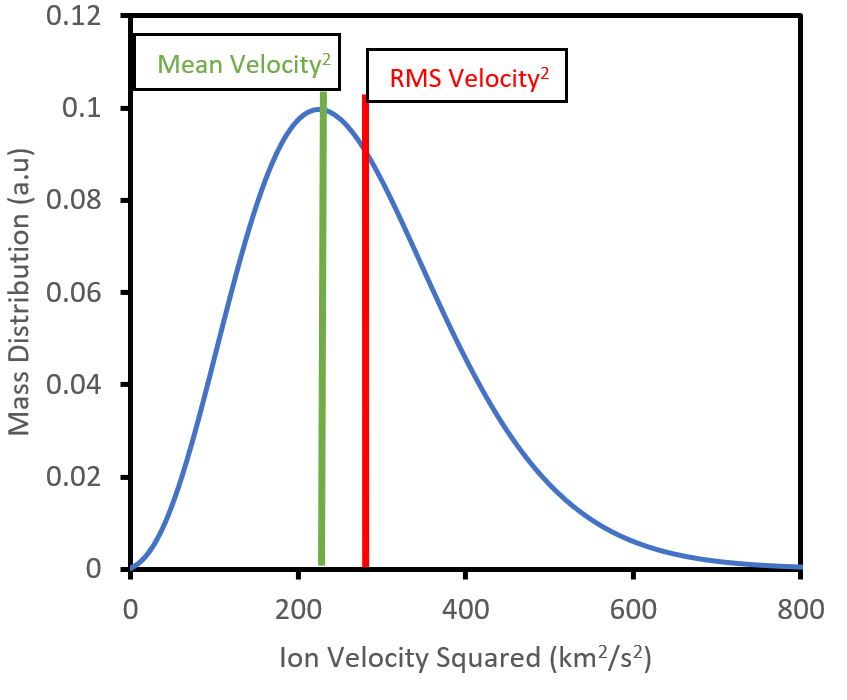}
\caption{Gaussian velocity distribution with both mean velocity squared and rms velocity squared}
\label{fig:RMSVelocitySquared}
\end{figure}

\begin{equation} \label{eq:KineticPowerNoMod}
{P_{Kinetic}} = \frac{\int \dot{m} {v}^2}{2}
\end{equation}

\begin{equation} \label{eq:ThrustPowerNoMod}
{P_{Thrust}} = \frac{{T}^2}{2 {\dot{m}_{tot}}} = \frac{(\int \dot{m} {v})^2}{2 \int \dot{m}}
\end{equation}

In any velocity distribution the RMS value is always greater than or equal to the mean. Proof of this can be found by the Schwartz Inequality, which for the thrust and kinetic power case can be seen by multiplying each side by the total mass flow.

\begin{equation} \label{eq:SchwarzInequality}
\left(\int \dot{m} {v}\right)^2 \leq \int \dot{m} {v}^2 \int \dot{m}
\end{equation}
meaning

\begin{equation} \label{eq:ThrusttoKineticInequality}
{P_{Thrust}} \leq {P_{Kinetic}}
\end{equation}
Equality between these two terms can be set by assigning an ion power ratio efficiency factor
\begin{equation} \label{eq:PowerUtilization}
\eta_{ipr} = \frac{P_{Thrust}}{P_{Kinetic}}
\end{equation}

The physical meaning behind the ion power ratio is the imbalance in kinetic energy fast and slow moving particles take compared to the thrust they provide, which results in an inefficiency of the transformed kinetic power. For a given mass flow rate and propellant utilization (i.e. ion flow), a lower input power is required to achieve the targeted thrust with mono energetic ions with the velocity $v_{jet}$ than with ions having a velocity distribution function including ions with $v_i> v_{jet}$ and $v_i<v_{jet}$. It implies that to achieve the same thrust, the presence of slow ions ($v_i<v_{jet}$) will need to be compensated by faster ions. The generation of these faster ions will require more power than would be needed for mono-energetic ions to produce the same thrust. As a result, a wider velocity distribution will have a lower ion power ratio and lower efficiency. Velocity distributions are common in plasma thrusters - accelerated ions often have a low-velocity tail due to ions born downstream of the ionization region. Efficiency is then decreasing not only by the lowered mean exhaust velocity (and so voltage efficiency), but also by an inefficient transformation of kinetic power to thrust power by the ion power ratio. It is important to distinguish the fraction of the input electric which goes to the kinetic power of the ions and the fraction of the input power which goes directly to the thrust generation. The former is defined by the current utilization efficiency times the voltage efficiency. The latter is by the inclusion of this term which considers the velocity distribution, where the presence of slow and fast ions decreases the portion of kinetic power that is converted to thrust power.

This is concerning for oscillating thrusters as it is inherent in the device that there is a distribution in velocity because ion velocity is changing with time. Due to the smaller timescale of velocity oscillations compared to the thrust operation, the varying velocities (with time) are grouped into a single distribution of which the ion power ratio applies. This represents an inherent decrease in efficiency of such devices. The degree to which this is a problem for sinusoidal oscillations is analyzed in Section 5.B. For demonstration purposes, the case of no oscillations and two species (ions and neutrals) with single velocities will be considered first, where it can be shown that the propellant utilization falls out of the ion power ratio. It should be noted the ion power ratio is similar to the squared inverse of the "form factor" used in electrical engineering. The form factor is the ratio of RMS signal to mean signal, where in this case the form factor is of the velocity distribution.

\subsection{No Oscillations - two species}

When deriving the total efficiency the ion power ratio is either ignored by assuming a single species of propellant, which essentially assigns the velocity distribution as a delta function and collapses the integrals to provide equality between thrust and kinetic power, or it is assumed there are a discrete number of species. Usually this is taken to be a singly charged ion and a neutral species, which turns the ion power ratio into the propellant utilization. This can be seen by taking the above ion power ratio and assuming the velocity distribution is the sum of a delta function for each species. Taking Eq. \ref{eq:SchwarzInequality} with $\dot{m} = \dot{m}_n\delta(v_n) + \dot{m}_i \delta(v_i)$ where $v_n$ and $v_i$ are the neutral and ion velocity respectively and $ \dot{m}_n$ and $ \dot{m}_i$ are the neutral and ion mass flow respectively:

\begin{equation} \label{eq:SchwarztoProp1}
\begin{aligned}
\left(\int \dot{m} {v}\right)^ 2&= \left(\int (\dot{m}_n\delta(v_n) + \dot{m}_i \delta(v_i) ){v} dv\right)^2\\
&= (\dot{m}_n v_n + \dot{m}_i v_i)^2 \\
&= (\dot{m}_n v_n)^2 + (\dot{m}_i v_i)^2 + 2\dot{m}_i v_i \dot{m}_n v_n
\end{aligned}
\end{equation}

\begin{equation} \label{eq:SchwarztoProp2}
\begin{aligned}
\int \dot{m} {v}^2 \int \dot{m} &= \int \left(\dot{m}_n\delta(v_n) + \dot{m}_i \delta(v_i) \right){v}^2 dv \\
&\times \int \left(\dot{m}_n\delta(v_n) + \dot{m}_i \delta(v_i) \right)dv\\
&= (\dot{m}_n + \dot{m}_i)\left(\dot{m}_n v_n^2 + \dot{m}_i v_i^2  \right)
\end{aligned}
\end{equation}
where the inequality is solved by the propellant utilization form of the ion power ratio by assuming $v_i \gg v_n$ and plugging Eq. \ref{eq:SchwarztoProp1} and \ref{eq:SchwarztoProp2} into Eq. \ref{eq:PowerUtilization}.

\begin{equation} \label{eq:SchwarztoProp3}
\begin{aligned}
\eta_{prop} &= \frac{(\dot{m}_n v_n)^2 + (\dot{m}_i v_i)^2 + 2\dot{m}_i v_i \dot{m}_n v_n}{(\dot{m}_n + \dot{m}_i)\left(\dot{m}_n v_n^2 + \dot{m}_i v_i^2  \right)}\\
&\approx \frac{(\dot{m}_i v_i)^2}{(\dot{m}_n + \dot{m}_i)\left( \dot{m}_i v_i^2  \right)}\\
&\approx \frac{\dot{m}_i \dot{m}_i v_i^2}{(\dot{m}_n + \dot{m}_i) \dot{m}_i v_i^2}\\
&\approx \frac{\dot{m}_i}{\dot{m}_i+\dot{m}_n}
\end{aligned}
\end{equation}
Including a time dependence on the mass flow (or ion current) and velocity (or ion energy) precludes one from using the propellant utilization form of the ion power ratio or collapsing the integrals by delta functions. Instead the full integral must be solved.

\subsection{Oscillations - single species}

As a simplification only a single species of ions with a single velocity at any point in time is considered here. Kinetic power of the oscillating ion flow is the product of the ion current and the ion energy:

\begin{equation} \label{eq:KineticPowerMod}
\begin{aligned}
{P_{Kinetic}} &=
    \frac{1}{2 \pi} \int_{0}^{2 \pi} (I_{im} + I_{ia}\sin{(\theta + \phi_{i})})(V_{im} + V_{ia}\sin{\theta}) d\theta\\
    & = I_{im} V_{im}  + I_{ia} V_{ia}\cos{(\phi_{i})}/2
\end{aligned}
\end{equation}
The thrust power from a thruster with oscillations remains the same form as the no-oscillation version, where the square of the time-averaged thrust is divided by the time averaged mass flow. This is due to the fact that the spacecraft will experience the time-averaged thrust as it travels. It is in these time-averages that the nuance of the ion power ratio is found. One intuitive result revealing the ion power ratio can be seen when the phase angle $\phi_{i} = 90^{\circ}$. At high ion energy oscillations $V_{ia}$ the thrust decreases (see Fig. \ref{fig:phaseenergydependence}), while the kinetic power stays constant (see Eq. \ref{eq:KineticPowerMod}). Thus, the input electric power is being transferred into kinetic power of the ions that is not resulting in thrust power. The expression for $\eta_{ipr}$ can be very involved, depending on the order of expansion of the thrust. For this paper, we shall only expand up to the second order. Taking thrust power from Eq. \ref{eq:ThrustPowerMod}, writing $\bar i = I_{ia}/I_{im}$ and through some simplification:

\begin{equation} \label{eq:OscUtil1}
\begin{aligned}
{\eta_{ipr}} &= \frac{P_{Thrust}}{P_{Kinetic}}\\
&= \frac{\frac{(A + B)^2 }{4 \pi^2 I_{im}}}{I_{im} V_{im}  + I_{ia} V_{ia}\cos{(\phi_{i})}/2}\\
&= \frac{ \left(1-\frac{\bar \upsilon^2}{16}+\bar i\,\bar \upsilon\cos{(\phi_{i})}/4\right)^2 }{1+\bar i\,\bar \upsilon\cos(\phi_{i}) / 2}
\end{aligned}
\end{equation}

\begin{figure}
\centering
\includegraphics[width=0.5\textwidth]{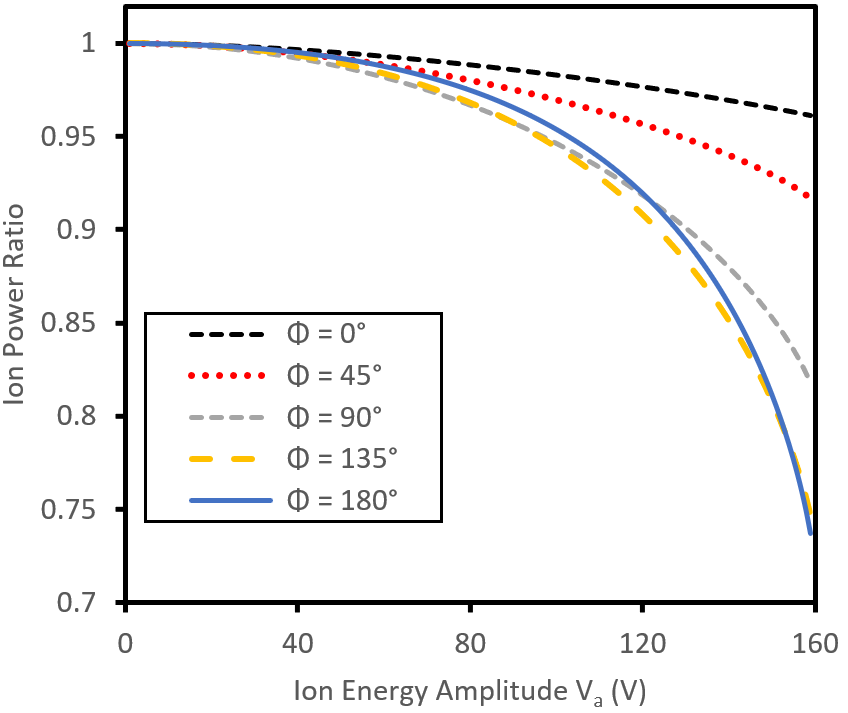}
\caption{Numerically calculated ion power ratio vs Ion Energy Amplitude over a range of phase angles for sinusoidal modulation. Thruster parameters: $V_{im} = 160 V$, $I_{im}=I_{ia}=0.3 A$}
\label{fig:oscutilfig}
\end{figure}

Similar solutions can be found for other orders of expansion. Fig. \ref{fig:oscutilfig} shows that ion power ratio decreases as low as 73\%, which represents nearly 30\% of kinetic power not contributing to thrust. This is the worst case scenario that exists for a thruster when the ion energy amplitude is equal to the ion mean energy. However, by controlling the phase of the ion energy and ion current it is possible to increase the ion power ratio to 95\%. This highlights the importance of ensuring the phasing of an oscillating thruster is in the optimal regime.

\section{Conclusions}
The main results and following conclusions have implications for plasma thrusters operating with natural discharge oscillations and for thrusters operating with externally driven oscillations. It is shown that the thrust can be increased with oscillations of the ion energy and the ion current. The maximum thrust is achieved when the two are in phase and oscillating with large amplitudes. A method to determine maximum thrust was shown for the out-of-phase case. It was shown that for a thruster with oscillating input voltage and current, performance is highest when the discharge current and the discharge voltage are out of phase and when ion current and ion energy are in phase. For sinusoidal oscillations, the thrust was found to potentially increase by 20\% or decrease up to 40\%.

Because the plasma oscillations can induce time variations of the ion velocity distribution function, we also analyzed the effect of the VDF on the ion power ratio; a generalized form of propellant utilization. The ion power ratio represents the portion of kinetic power that is transformed into the thrust power, which can be decreased by a wide velocity distribution of the exhaust. This revealed an inefficiency that can significantly decrease the performance of a thruster: for sinusoidal oscillations the ion power ratio can be as low as 73\%. By adjusting the phase between ion current and energy, however, this inefficiency can be nearly completely nullified. While the presented analysis was conducted for a specific waveform, the same approach can be taken for any arbitrary waveform. Future work may reveal waveforms with greater gains in thrust and thruster efficiency.

%\begin{appendices}
\section*{Appendix}

\subsection{Square Wave Thrust Equations}

\begin{figure}
\centering
\includegraphics[width=0.45\textwidth]{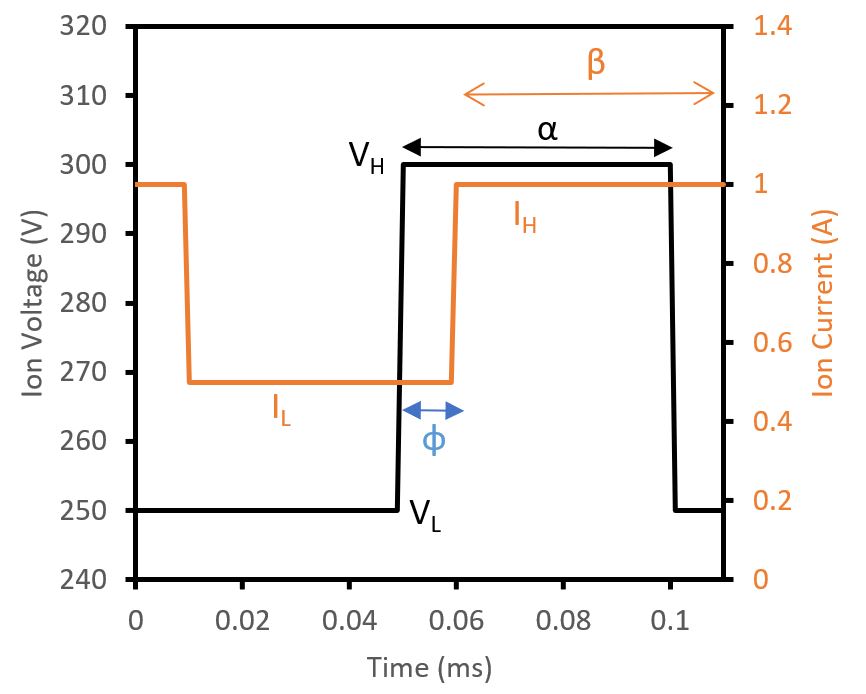}
\caption{Ion Voltage and Ion Current with Time at 10kHz Pulsed Operation}
\label{fig:highfreqtimes}
\end{figure}

\begin{equation} \label{eq:squarepulse}
    T= \sqrt{\frac{2M}{e}}
\begin{cases}
\begin{aligned}
&I_L\left( \sqrt{V_L}(1-\alpha)+\sqrt{V_H}(\alpha-\beta)\right)\\ &+I_H\sqrt{V_H}\beta\\
\end{aligned}
    ,& \text{if } \alpha\geq \phi_{i}+\beta\vspace{2.5mm}\\
\begin{aligned}
&  I_L\left(\sqrt{V_L}(1-\beta-\phi_{i})+ \sqrt{V_H}\phi_{i}\right)\\
&+ I_H\left(\sqrt{V_L}(\phi_{i}+\beta-\alpha) + \sqrt{V_H}(\alpha-\phi_{i})\right)\\
\end{aligned}
    ,& \text{if }  \phi_{i}+\beta \geq \alpha \geq \phi_{i}\vspace{2.5mm}\\
\begin{aligned}
& I_L( \sqrt{V_L}(\phi_{i}-\alpha)+\sqrt{V_H}(1+\alpha-\beta-\phi_{i}))\\
&+ I_H\left(\sqrt{V_L}(1-\phi_{i})+ \sqrt{V_H}(\phi_{i}+\beta-1)\right)\\
\end{aligned}
    ,& \text{if } \alpha+1-\beta \geq \phi_{i} \geq \alpha\vspace{2.5mm}\\
\begin{aligned}
& I_L\left( \sqrt{V_L}(\beta-1)\right)\\ 
&+I_H\left(\sqrt{V_L}(\beta-\alpha)+ \sqrt{V_H}(\alpha)\right)\\
\end{aligned}
    ,& \text{if } \phi_{i}+\beta-1 \geq \alpha\vspace{2.5mm}\\
\begin{aligned}
& I_L\left( \sqrt{V_H}(\beta-1)\right)\\ 
&+I_H\left(\sqrt{V_L}(1-\alpha)+ \sqrt{V_H}(\alpha+\beta-1)\right)\\
\end{aligned}
    ,& \text{if } \beta \geq \alpha \geq \phi_{i}\vspace{2.5mm}\\
\begin{aligned}
& I_L\left(\sqrt{V_L}(1-\alpha-\beta) \sqrt{V_H}(\alpha)\right)\\ 
&+I_H\left(\sqrt{V_L}(\beta)+ \sqrt{V_H}(\alpha)\right)
\end{aligned}
    ,& \text{if } 1-\beta \geq \phi_{i} \geq \alpha\\
\end{cases}
\end{equation}

For a square wave oscillation as shown in Fig. \ref{fig:highfreqtimes} these solutions are shown in Eq. \ref{eq:squarepulse} where $\alpha$ is the duty cycle of the ion modulation, $\beta$ is the duty cycle for the ion current modulation, and $\phi_{i}$ is the dimensionless phase shift between the two parameters. All parameters are shown graphically in Fig. \ref{fig:highfreqtimes}, and the particular duty cycle/phase-shift combination observed in the figure is described by the second case in Eq \ref{eq:squarepulse}. 

\subsection{Full Derivation of Thrust - Oscillations}
Thrust for a thruster with oscillations is found by taking the time-average of the instantaneous thrust over the oscillations, where the instantaneous thrust has the form  $T =  \dot{m} v_{ex}$. Here we are assuming a single species of exhaust.
\begin{equation} \label{eq:ThrustMod1}
\begin{aligned}
    {T_{mod}} &= \frac{1}{\tau} \int_{0}^{\tau} \dot{m}(t) v_{ex}(t) dt \\        
    &= \frac{1}{\tau} \int_{0}^{\tau} \dot{m}(t) \sqrt{2\, K(t) / M} dt
  \end{aligned}
\end{equation}
For purposes of equating the thrust and kinetic power later for electric propulsion devices, the mass flow and exhaust energy will be written as current ($I_{i} = \dot{m} e / M$) and voltage ($V_{i}=K/e$)respectively. Note that this analysis is not restricted to ion propellant, as these are only separated from mass flow and energy by a constant.

\begin{equation} \label{eq:ThrustMod2}
\begin{aligned}
    {T_{mod}} &= \frac{1}{\tau} \int_{0}^{\tau} \frac{I_i(t) M}{e} \sqrt{\frac{2 e V_i(t)}{M}} dt\\        
    &= \frac{1}{2 \pi} \int_{0}^{2 \pi} \frac{I_i(\theta) M}{e} \sqrt{\frac{2 e V_i(\theta)}{M}} d\theta\\
    &=  \sqrt{\frac{2 M}{e}}\frac{1}{2 \pi} \int_{0}^{2 \pi} I_i(\theta) \sqrt{V_i(\theta)} d\theta
  \end{aligned}
\end{equation}
The form of the waveform of both the mass flow and the exhaust energy is important. Here we will assume each is an offset sinusoid as shown in Fig. \ref{fig:highfreqsine}. That is a sinusoid which has an offset larger than the amplitude such that it is never negative. To account for a possible phase shift between the energy and mass flow, which measurements of a modulated Hall Thruster have shown to exist, a phase angle $\phi_{i}$ is included.
\begin{equation} \label{eq:ThrustMod3}
\begin{gathered}
    {T_{mod}} = \\
    \sqrt{\frac{2 M}{e}}\frac{1}{2 \pi} \int_{0}^{2 \pi} (I_{im} + I_{ia}\sin{(\theta + \phi_{i})}) \sqrt{(V_{im} + V_{ia}\sin{\theta})} d\theta\\
    =  \sqrt{\frac{2 M}{e}}\frac{1}{2 \pi} \left( \int_{0}^{2 \pi} I_{im} \sqrt{(V_{im} + V_{ia}\sin{\theta})} d\theta\right. \\
    +  \int_{0}^{2 \pi}I_{ia}\sin{\theta}\cos{\phi_{i}} \sqrt{(V_{im} + V_{ia}\sin{\theta})} d\theta\\
    +  \left.\int_{0}^{2 \pi}I_{ia}\cos{\theta}\sin{\phi_{i}} \sqrt{(V_{im} + V_{ia}\sin{\theta})} d\theta\right)\\
    = \sqrt{\frac{2 M}{e}}\frac{1}{2 \pi}(A + B + C)
  \end{gathered}
\end{equation}

Each integral (A, B, and C) will be solved separately.

\begin{equation} \label{eq:IntegralA1}
\begin{aligned}
    A &=  \int_{0}^{2 \pi} I_{im} \sqrt{(V_{im} + V_{ia}\sin{\theta})} d\theta\\
    &= I_{im} \sqrt{V_{im}} \int_{0}^{2 \pi}\sqrt{1 + \frac{V_{ia}}{V_{im}}\sin{\theta}} d\theta\\
  \end{aligned}
\end{equation}
Eq. \ref{eq:IntegralA1} can be reduced to a form involving elliptic integrals of the second kind, E(k). Taking $\bar \upsilon = V_{ia}/V_{im}$ as our independent variable:
\begin{equation} \label{eq:IntegralA2}
\begin{aligned}
    &A(\bar \upsilon) =  \\
    &2 I_{im} \sqrt{V_{im}} \left(\sqrt{1-\bar \upsilon} E \left(\frac{2 \bar \upsilon}{\bar \upsilon-1}\right) + \sqrt{1+\bar \upsilon} E \left(\frac{2 \bar \upsilon}{1+\bar \upsilon}\right)\right)
  \end{aligned}
\end{equation}
The complete elliptic integral of the second kind can be expressed by the power series
\begin{equation} \label{eq:EllipticIntegralE}
    E(k) = \frac{\pi}{2} \sum_{n=0}^{\infty}\left(\frac{(2n)!}{2^{2n}(n!)^2}\right)^2 \frac{k^{2n}}{1-2n}
\end{equation}
Integral A can then be simplified to:
\begin{equation} \label{eq:IntegralA3}
\begin{aligned}
    A(\bar \upsilon) &= 2 \pi I_{im} \sqrt{V_{im}} \left( 1 - \frac{\bar \upsilon^2}{16} - \frac{15 \bar \upsilon^4}{1024} - \frac{105 \bar \upsilon^6}{16384} - ...   \right)
  \end{aligned}
\end{equation}

A similar approach is taken to find integral B
\begin{equation} \label{eq:IntegralB1}
\begin{aligned}
    B &=  \int_{0}^{2 \pi}I_{ia}\sin{\theta}\cos{\phi_{i}} \sqrt{V_{im} + V_{ia}\sin{\theta}} d\theta\\
    &=  I_{ia} \cos{\phi_{i}}\sqrt{V_{im}} \int_{0}^{2 \pi}\sin{\theta} \sqrt{1 + \frac{V_{ia}}{V_{im}}\sin{\theta}} d\theta\\
  \end{aligned}
\end{equation}
Again taking $\bar \upsilon = V_{ia}/V_{im}$ as our independent variable, the solution of Eq. \ref{eq:IntegralB1} can be solved into a form involving the elliptic integrals of both the first $K(k)$ and second kind $E(k)$.

\begin{equation} \label{eq:IntegralB2}
\begin{aligned}
    B(\bar \upsilon) &=  \frac{2 I_{ia} \cos{\phi_{i}}\sqrt{V_{im}}}{3 \bar \upsilon}\biggl( \\
    & \sqrt{1-\bar \upsilon} \left( E \left(\frac{2 \bar \upsilon}{\bar \upsilon-1}\right)
    - \left(\bar \upsilon+1\right)K\left(\frac{2 \bar \upsilon}{\bar \upsilon-1}\right)\right)\\
    &+ \sqrt{\bar \upsilon+1} \left( E \left(\frac{2 \bar \upsilon}{\bar \upsilon+1}\right)
    - \left(\bar \upsilon-1\right)K\left(\frac{2 \bar \upsilon}{\bar \upsilon+1}\right)\right)\biggr)
  \end{aligned}
\end{equation}
Solving for the power series of Eq. \ref{eq:IntegralB2} provides a simplified form which quickly converges.

\begin{equation} \label{eq:EllipticIntegralK}
    K(k) = \frac{\pi}{2} \sum_{n=0}^{\infty}\left(\frac{(2n)!}{2^{2n}(n!)^2}\right)^2 k^{2n}
\end{equation}

\begin{equation} \label{eq:IntegralB3}
\begin{aligned}
    B(\bar \upsilon) &=  2 \pi I_{ia} \cos{\phi_{i}}\sqrt{V_{im}}\left( \frac{\bar \upsilon}{4} + \frac{3 \bar \upsilon^3}{126} + \frac{35 \bar \upsilon^5}{4096}+... \right)
  \end{aligned}
\end{equation}

Integral C is simply zero, which can be shown by method of u substitution. Taking $u = V_{im} + V_{ia} \sin{\theta}$:
\begin{equation} \label{eq:IntegralC1}
\begin{aligned}
    C &=  \int_{0}^{2 \pi}I_{ia}\cos{\theta}\sin{\phi_{i}} \sqrt{V_{im} + V_{ia}\sin{\theta}} d\theta\\
    &=  I_{ia} \sin{\phi_{i}} \int_{0}^{2 \pi}\cos{\theta} \sqrt{V_{im} + V_{ia}\sin{\theta}} d\theta \\
    &=  I_{ia} \sin{\phi_{i}} \int_{\theta = 0}^{\theta = 2 \pi}\frac{\cos{\theta} \sqrt{u}}{V_{ia}\cos{\theta}} du \\
    &= \frac{I_{ia} \sin{\phi_{i}}}{V_{ia}} \int_{\theta = 0}^{\theta = 2 \pi}\sqrt{u} du \\
    &= \frac{2 I_{ia} \sin{\phi_{i}}}{3 V_{ia}} \left( (V_{im} + V_{ia} \sin{2 \pi})^{3/2} - (V_{im} + V_{ia} \sin{0})^{3/2}\right)\\
    &= 0 
  \end{aligned}
\end{equation}

For a thruster with an oscillation in the mass flow and energy of an offset-sinusoid form, the thrust can then be written as:

\begin{equation} \label{eq:ThrustModFinal}
\begin{gathered}
 T_{mod} =  \sqrt{\frac{2 M V_{im}}{e}}\left( I_{im}\left(1- \sum_{n=1}^{\infty} a_{2n}\, \bar \upsilon^{2n}\right)\right.\\
 \left.+ I_{ia}\cos\phi_{i}\sum_{n=1}^{\infty} a_{(2n-1)}\,\bar \upsilon^{(2n-1)}\right)
  \end{gathered}
\end{equation}
where the coefficients $a_n$ can be found through the power series of the elliptic integrals. The first 6 terms are shown in Eq. \ref{eq:IntegralA3} and Eq. \ref{eq:IntegralB3}. When there is no oscillation in ion energy or when the expansion is taken to the zeroth order, Eq. \ref{eq:ThrustModFinal} reduces to $ T =  \sqrt{\frac{2 M V_{im}}{e}} I_{im} = v_{jet} \dot{m}_{tot}$

\begin{figure}

\centering
\includegraphics[width=0.5\textwidth]{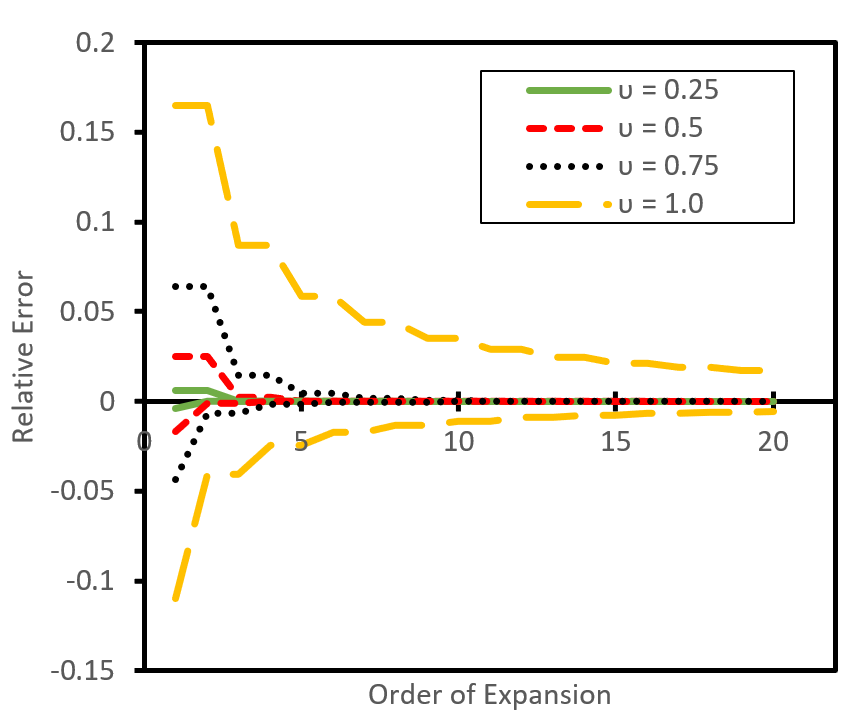}
\caption{Error of integral A (positive) and B (negative) vs the expansion order over a range of $V_{ia}/V_{im}$.} \label{fig:errorintegrals}
\end{figure}

While it is possible to solve the modulated thrust for expansion orders to the nth degree, the series quickly converges, particularly for lower $\bar \upsilon=V_{ia} / V_{im}$. The error of the series expansion for integral A and integral B are shown in Fig. \ref{fig:errorintegrals}. Error is defined here as the difference between the series expansion and the numerically calculated divided by the numerically calculated. Fig. \ref{fig:errorintegrals} show the error below 2\% at 3rd order for $\bar \upsilon$ $<$ 0.5. For thrusters with much fuller oscillations with $\bar \upsilon$ $\sim$1, error on thrust is less than 5\% with 6th order expansion. Several useful forms of the thrust follow:

If the ion energy oscillation amplitude is less than half the mean ion energy the thrust can be expanded to the first order with error below 4\%:

\begin{equation}\label{eq:practicalthrust1}
    T_{mod} \approx  \sqrt{\frac{2 M V_{im}}{e}}\left( I_{im} + I_{ia}\cos\phi_{i}\frac{V_{ia}}{4 V_{im}}\right)
\end{equation}
If the ion energy oscillation amplitude is equal to the mean ion energy (full pulse) an exact expression can be found:
\begin{equation}\label{eq:practicalthrust2}
    T_{mod} =  \frac{4}{3\pi}\sqrt{\frac{M V_{im}}{e}}\left( 3 I_{im}\ + I_{ia}\cos\phi_{i}\right)
\end{equation}

\subsection{Maximum Thrust Derivation}

To derive voltage amplitude that will provide the maximum thrust, we take our derived expression of thrust (Eq. \ref{eq:ThrustModFinalSec}) and take the derivative with respect to $\bar \upsilon$.

\begin{equation} \label{eq:ThrustModDeriv1}
\begin{gathered}
 \frac{dT_{mod}}{d\bar \upsilon} =  \sqrt{\frac{2 M V_{im}}{e}}\left( I_{im}\left(- \sum_{n=1}^{\infty} 2n\,a_{2n}\, \bar \upsilon^{2n-1}\right)\right.\\
 \left.+ I_{ia}\cos\phi_{i}\sum_{n=1}^{\infty} (2n-1)\,a_{(2n-1)}\,\bar \upsilon^{(2n-2)}\right)
  \end{gathered}
\end{equation}
We then set the left-hand side of Eq. \ref{eq:ThrustModDeriv1} to zero and simplify.

\begin{equation} \label{eq:ThrustModDeriv2}
\begin{gathered}
 0 =  \sqrt{\frac{2 M V_{im}}{e}}\left( I_{im}\left(- \sum_{n=1}^{\infty} 2n\,a_{2n}\, \bar \upsilon^{2n-1}\right)\right.\\
 +\left. I_{ia}\cos\phi_{i}\sum_{n=1}^{\infty} (2n-1)\,a_{(2n-1)}\,\bar \upsilon^{(2n-2)}\right)
  \end{gathered}
\end{equation}

\begin{equation} \label{eq:ThrustModDeriv3}
\begin{gathered}
 I_{im} \sum_{n=1}^{\infty} 2n\,a_{2n}\, \bar \upsilon^{2n-1} = I_{ia}\cos\phi_{i}\sum_{n=1}^{\infty} (2n-1)\,a_{(2n-1)}\,\bar \upsilon^{(2n-2)}
  \end{gathered}
\end{equation}

\begin{equation} \label{eq:maxthrustder2}
\frac{\sum_{n=1}^{\infty} 2n\,a_{2n}\, \bar \upsilon^{2n-1}}{\sum_{n=1}^{\infty} (2n-1)\,a_{(2n-1)}\,\bar \upsilon^{(2n-2)}} =  \frac{I_{ia}\cos{\phi_{i}}}{I_{im}} 
\end{equation}
This expression can then be numerically solved for $\bar \upsilon$, as was performed in Fig. \ref{fig:maxthrustva}.

\subsection{Thrust Expansion Coefficients}\label{subsec:coeffs}

The coefficients $a_n$ for the series expansion in Eq. \ref{eq:ThrustModFinal} are shown up to 12th order. These coefficients can be calculated through the series expansion of integrals A and B.

\begin{equation}
\begin{aligned}
a_1 &= \frac{1}{4} = 0.2500\\
a_2 &= \frac{1}{16} = 0.0625\\
a_3 &= \frac{3}{128} = 0.0234\\
a_4 &= \frac{15}{1024} = 0.0146\\
a_5 &= \frac{35}{4096} = 0.0085\\
a_6 &= \frac{105}{16384} = 0.0064\\
a_7 &= \frac{1155}{262144} = 0.0044\\
a_8 &= \frac{15015}{4194304} = 0.0036\\
a_9 &= \frac{45045}{16777216} = 0.0027\\
a_{10} &= \frac{153153}{67108864} = 0.0023 \\
a_{11} &= \frac{969969}{536870912} = 0.0018\\
a_{12} &= \frac{6789783}{4294967296} = 0.0016\\
\end{aligned}
\end{equation}

%

%
%\end{appendices}
\section*{Funding Sources}
This work was supported by the Air Force Office of Scientific Research.

\section*{Acknowledgments}
The authors would like to thank Prof. Andrei Smolyakov for fruitful discussions. 

\bibliography{main}% Produces the bibliography via BibTeX.

\end{document}